\documentclass[aps,prl,10pt,superscriptaddress,twocolumn,footinbib,pdfstartview=FitH,colorlinks=true,citecolor=blue,linkcolor=blue,urlcolor=blue]{revtex4-2}
\usepackage[cm]{fullpage}
\usepackage{orcidlink}
\usepackage{microtype}
\usepackage{amsfonts,amsmath,amssymb,bm,mathtools,physics}
\usepackage{graphicx,xcolor}

\def\sss{\scriptscriptstyle}

\begin{document}

\preprint{APS/123-QED}

\author{Myrann Baker-Rasooli~\orcidlink{0000-0003-0969-6705}}
\affiliation{Laboratoire Kastler Brossel, Sorbonne Universit\'e, CNRS,
ENS-PSL Research University, Coll\`ege de France, 4 Place Jussieu, 75005 Paris, France}

\author{Nathan du Toit~\orcidlink{0009-0000-4988-4722}}
\affiliation{Université Paris-Saclay, CNRS, LPTMS, 91405 Orsay, France}

\author{Nicolas Pavloff~\orcidlink{0000-0001-8814-7848}}
\affiliation{Université Paris-Saclay, CNRS, LPTMS, 91405 Orsay, France}
\affiliation{Institut Universitaire de France (IUF)}

\author{Quentin Glorieux~\orcidlink{0000-0003-0903-0233}}
\affiliation{Laboratoire Kastler Brossel, Sorbonne Universit\'e, CNRS,
ENS-PSL Research University, Coll\`ege de France, 4 Place Jussieu, 75005 Paris, France}
\affiliation{Institut Universitaire de France (IUF)}

\title{Vortex leapfrogging and superfluid dissipation mechanisms in a  fluid of light}

\begin{abstract}

We report the experimental observation of vortex leapfrogging in a two-dimensional fluid of light. 
By imprinting two vortex–antivortex pairs and tracking their real-time evolution through phase-resolved imaging, we observe a dynamics that is accurately described by a point-vortex model with an outward background flow.
By precisely controlling the initial vortex separation, we identify configurations in which leapfrogging breaks down and determine the corresponding dissipation mechanisms. The first originates from phase-slip events occurring at large injected velocities.
The second arises when the injection of multi-charged vortices leads to the formation of a dispersive shock 
wave which acts as a continuous source of phase slippage.
These mechanisms advance our understanding of vortex dynamics and dissipation in superfluids.

\end{abstract}

\maketitle

 Few fluid-dynamic phenomena are as visually striking as vortex leapfrogging, where a pair of vortex rings repeatedly pass through one another in an alternating sequence.
In classical fluids, vortex leapfrogging has long been a canonical problem of hydrodynamics~\cite{helmholtz1858,Thomson1868,meleshko2010}. Its quantum counterpart, although extensively explored in simulations~\cite{Wacks2014, Caplan2014, Freund2018,  Ikuta2019, Barenghi2021}, still lacks from experimental demonstration.
In two-dimensional quantum fluids, the most simple theoretical description of the leapfrogging mechanism is the Helmholtz point-vortex model~\cite{Helmholtz1867}, which captures the key features of vortex–vortex interactions in an inviscid, incompressible flow but neglects compressibility, finite vortex-core size, and dissipation. Because these neglected effects are often crucial in real quantum gases ~\cite{panico2025}, experimentally probing leapfrogging and its possible breakdown provides an important benchmark for superfluid hydrodynamics.

An important concept of superfluids (and superconductors) is phase coherence and the ability to sustain currents without dissipation. 
In such systems, the macroscopic wave function acts as an order parameter whose phase gradient sets the superfluid velocity. 
However, this metastable flow is known to decay via phase slips: As initially suggested by Anderson \cite{Anderson1966} localized events where the amplitude of the order parameter is zero and the phase jumps by $2\pi$, leading the superfluid to relax to a state with lower velocity~\cite{Moulder2012, Wright2013,Jendrzejewski2014}. 
In this context, the study of vortex leapfrogging provides an ideal playground to contribute to the microscopic description of the effective dissipation in an otherwise conservative dynamic, by observing the point-vortex model breaking down when vortices annihilate, merge, or nucleate in pairs~\cite{Kwon2021, congy2024, baker2025}.

In this Letter, we observe vortex leapfrogging in a two-dimensional quantum fluid of light and use it to study phase-slip–induced dissipation in superfluids. 
We show that, at early times, the vortex trajectories are well described by a point-vortex model in the presence of an outward background flow. 
We then identify the dissipation mechanisms responsible for the breakdown of leapfrogging.
The first is discrete phase slippage~\cite{Anderson1966}, in which vortices and other critical points are nucleated in a low-density, high-velocity region, leading to quantized $2\pi$ phase jumps and a reduction of the superfluid flow. 
The second arises in a strongly nonlinear regime, where the flow between two multicharged, counter-rotating vortex clusters becomes quasi-one-dimensional and generates a dispersive shock wave (DSW). 
In recent years,
DSWs have attracted considerable interest in Bose superfluids driven far from equilibrium,
especially in the context of quenches, collisions, or localized
density imbalances~\cite{hoefer2006,  Wan2007,  hoefer2008, Gang2017,
bienaime2021}. Here, we argue that the DSW acts as an 
amplification mechanism for the phase-slippage process and behaves as a continuous source of effective dissipation for the superfluid flow.

Paraxial fluids of light~\cite{glorieux2025paraxial} offer an ideal platform to investigate these phenomena. 
The propagation of light in a nonlinear medium is governed by an equation formally analogous to the Gross–Pitaevskii equation, enabling controlled studies of quantum hydrodynamics and vortex dynamics in an optical setting~\cite{fontaine2018observation,glorieux2023hot,PhysRevA.98.023825,michel2018superfluid,sitnik2022, azam2022vortex, baker2023turbulent, panico2023onset}. 
Crucially, the full complex order parameter i.e. both the intensity and the phase of the optical field can be measured, the latter providing direct access to the velocity potential of the flow~\cite{glorieux2025paraxial}. 
This capability allows high-resolution reconstruction of the full velocity field, including vortices and other critical points, and is therefore particularly well suited to understand the microscopic structure of dissipation mediated by phase slips and dispersive shock waves.\\

%%% FIG 1 %%%%
\begin{figure}[ht!]
    \centering
    \includegraphics[width=0.96\linewidth]{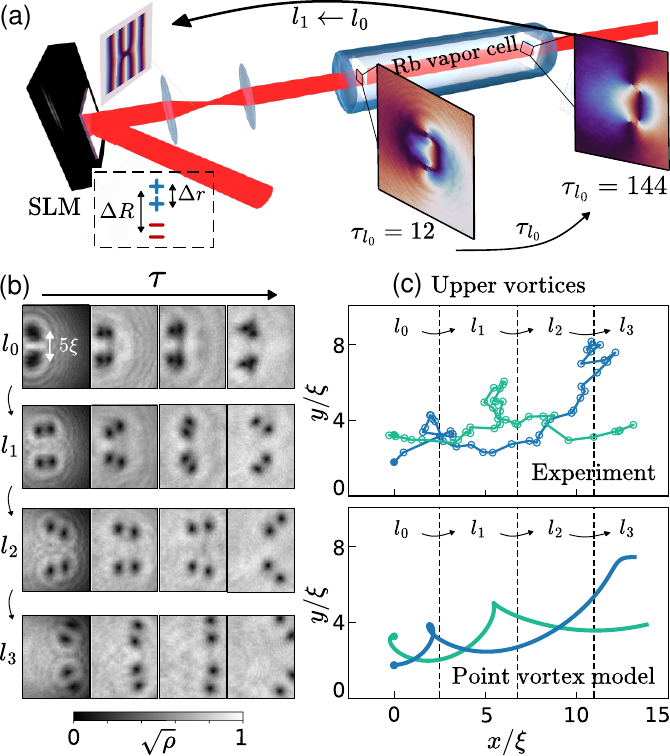}
    \caption{\textbf{Temporal evolution of vortex leapfrogging.} \textbf{(a)} Simplified setup. 
    A laser beam is sent on an SLM and imaged at the input of the 20~cm-long Rb vapor cell. 
    The phase modulation leads to the creation of two counter-rotating pairs vortices in the transverse plane. 
    The output plane of the nonlinear medium is imaged  after an interferometer (not shown).
     \textbf{(b)} Experimental images of the field amplitude.
     The rows $l_0$ to $l_3$ display the temporal evolution within each single loop. The last panel of row $l_n$ is recorded and re-injected as first panel of loop $l_{n+1}$.
     \textbf{(c)} Upper vortices trajectory (top) and point vortex model prediction (bottom) {for $\Delta {R} = 5\xi$ and $\Delta {r} = 1.5\xi$}.
     The dashed line indicates the $x$-coordinate of the vortex pair barycenter at each loop transition.}
    \label{fig:setup}
\end{figure}
%%% FIG 1 %%%%
A paraxial fluid of light consists of a monochromatic laser beam propagating through a non-linear medium. 
In the paraxial approximation, the propagation equation of the laser electric field envelope $\mathcal{E}$ is isomorphic to the Gross-Pitaevskii equation describing the temporal evolution of the wavefunction for a weakly interacting quantum gas.
Each transverse plane at fixed $z$ is then a temporal snapshot of the evolution for an ideal 2D system and the envelope $\mathcal{E}(\boldsymbol{r}, z )$ obeys the following equation: 
\begin{equation}
    \label{eq:NLSE}
    i \pdv{\mathcal{E}}{z} \  =  \left[ -\frac{1}{2 n_0 k_0} \boldsymbol{\nabla}_{\!\perp}^2 - \frac{k_0 \chi^{(3)}}{2 n_0} \, |\mathcal{E} |^2  \right]\mathcal{E},
\end{equation}
where $k_0$ is the wavevector, $n_0$ is the linear refractive index, the $\boldsymbol{\nabla}_{\!\perp}$ operator is defined as acting in the transverse $\boldsymbol{r}=(x,y)$ plane and the nonlinear term is proportional to  the laser intensity times the third-order susceptibility  $\chi^{(3)}$ at the laser frequency.
This cubic nonlinearity induces an effective photon-photon interaction which is set repulsive ($\chi^{(3)}<0$) to ensure a stable superfluid.
In the experiment, we create a fluid of light by propagating a 780~nm laser set close to resonance of the $^{87}$Rb D2 line  (detuned by $\sim -6$GHz of the $F=2\to F'$ transition) within a warm vapor cell ($T\sim 150^\circ$C and $L=20$~cm) of rubidium which acts as a nonlinear medium, see \cite{supplemental} for details. 

The light field within the non-linear medium is not directly accessible experimentally, yet temporal evolution may be retrieved using an adimensional form of Eq. (\ref{eq:NLSE}).
This is done by incorporating the interaction term into a rescaled variable $\tau=~{L}/{z_{\text{NL}}}$, where $z_{\text{NL}}=\left[-k_0 \chi^{(3)} |\mathcal{E} ({0, 0 })|^2/(2 n_0)\right]^{-1}$  is the characteristic nonlinear axial length and $L$ is the length of the non-linear cell \cite{bienaime2021}.
After re-scaling the transverse quantities ($\tilde{\boldsymbol{r}}=\boldsymbol{r}/{\xi}$, $\tilde{\boldsymbol{\nabla}}_{\perp} = {\xi}\boldsymbol{\nabla}_{\perp}$) by the transverse healing length ${\xi}= (z_\text{NL}/k_0)^{1/2}$, one obtains for $\psi(\tilde{\boldsymbol{r}}, \tau )=
\mathcal{E}(\boldsymbol{r}, z )/|\mathcal{E}({ 0, 0 })|$:
\begin{equation}
    i\frac{\partial\psi }{\partial \tau}=
    \left(-\frac{1}{2}
    \tilde{\boldsymbol{\nabla}}^2_{\perp}+{\mid}\psi{\mid}^2
    \right)\psi.
    \label{GPE_Adim}
\end{equation}
In this form, we see that the dynamics of the system can be studied by tuning the ratio $\tau=~{L}/{z_{\text{NL}}}$ which is monitored by the incoming laser intensity $|\mathcal{E} (0, 0 )|^2$.

As shown in Fig.~\ref{fig:setup}(a), the system is initialized by imprinting two vortex-antivortex pairs in the transverse plane of the laser beam using a Spatial Light Modulator (SLM) \cite{Arrizon07} (see details in End Matter). As illustrated on the inset, the distance 
 $\Delta{r}$
between vortices of the same vorticity
and the mean distance $\Delta{R}$
between  positive and negative vortices are fixed at the input of the medium.
Two vortices of positive sign are positioned in the upper part of the beam, and two others with negative sign are placed symmetrically in the lower part with $\Delta{r}/\xi=1.5$ and $\Delta{R}/\xi= 5$.
The complex order parameter, i.e. the electric field intensity and phase are recorded at the cell output using an off-axis interferometer \cite{PhaseUtils} and typical images are shown in Fig.~\ref{fig:setup}(a-b).

%%% FIG 2 %%%%
\begin{figure*}[!t]
    \centering
    \includegraphics[width=0.99\linewidth]{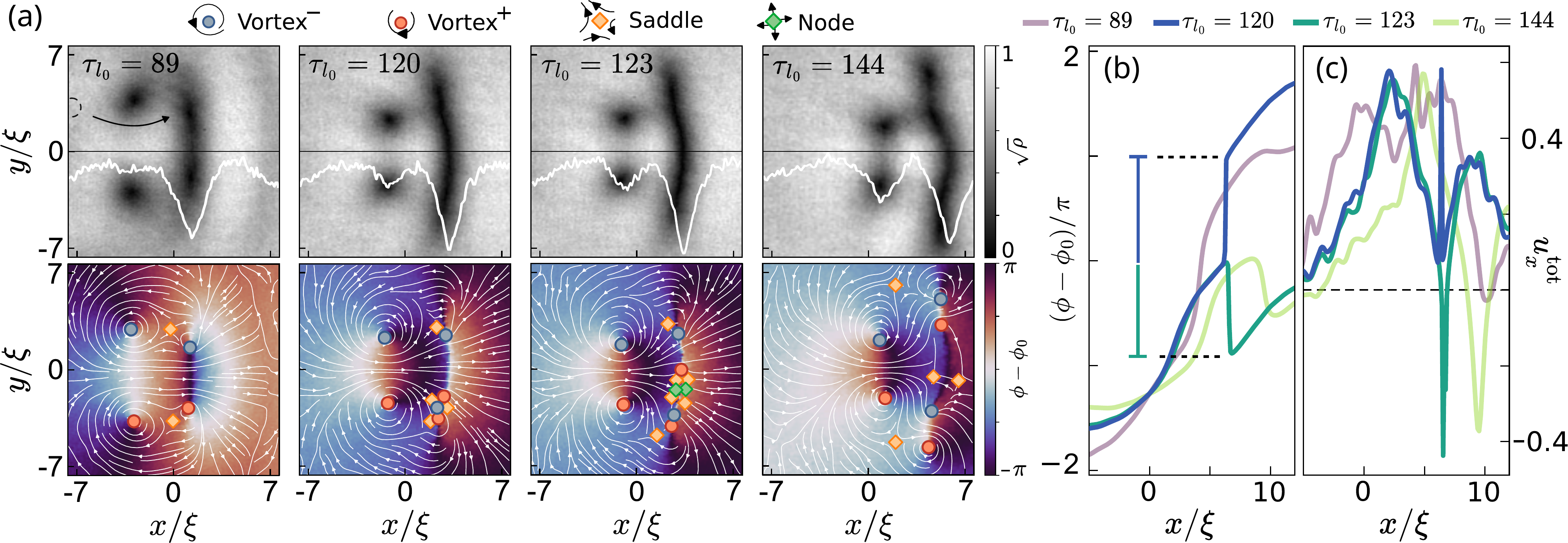}
    \caption{\textbf{Nucleation of phase slips.} Leapfrogging evolution in the first loop from left to right $\tau_{l_0}=89,120,125,144$ with vortex-antivortex pairs initially at positions $(0,\pm 1.75\xi)$ and $(0,\pm 3.25\xi)$.
     \textbf{(a)} - Top: experimental images of the field amplitude with the amplitude profile along the $x$-axis (white curve).
     Bottom: associated phase. The phase $\phi_0$ measured in the absence of any vortices is removed. The oriented white curves are streamlines.
     The topological charges are indicated by blue and red dots (vortices), orange diamonds (saddles) and green diamonds (nodes).
     % {\color{red}Bottom: associated equiphase lines. The phase slip mechanism is highlighted by the white lines.}
     \textbf{(b)} - Unwraped phase profile along the $x$-axis. 
     The $2\pi$ phase slip is clearly observed between $\tau=120$ and $\tau=123$.
     \textbf{(c)} - Total velocity field along the $x$-axis. 
     A large positive extremun at $\tau=120$ is immediately followed by a negative minimum at $\tau=123$.}
    \label{fig:vacuum}
\end{figure*}
%%% FIG 2 %%%%

The effective time is increased by ramping the laser power, thereby reducing $z_{\text{NL}}$ and adjusting in parallel the initial conditions to maintain the initial transverse quantities $\Delta{r}/\xi$ and $\Delta{R}/\xi$ constant.
However, the maximum achievable $\tau$ is $\sim 140$, since both $L$ and $z_{\text{NL}}$ are limited by the finite size of the cell and the laser intensity, respectively.
To overcome this constraint, we artificially extend the cell length through a feedback loop mechanism \cite{ferreira2024exploring}.
By measuring the output phase at the final time $\tau\simeq 140$, we detect the positions and charges of the vortices. 
We then reconstruct the analytical phase and density map of the four vortices from their positions and we reinject this wavefunction \cite{bradley2012energy} at the input as shown in  Fig.~\ref{fig:setup}.
This approach avoids  the accumulation of measurement noise and enables to observe more than a full cycle of vortex leapfrogging, corresponding to four complete feedback loops.

Fig.~\ref{fig:setup}(b) shows density snapshots at times $\tau_{l_i}$ (column) during each of the loops $l_i$ (lines), illustrating the vortices moving toward positive $x$ in the transverse plane.
Fig.~\ref{fig:setup}(c) shows the trajectories of the two vortices located in the upper half plane ($y>0$), over the four loops.
The top panel presents the experimental results, where each point is averaged over five realizations. The dashed lines indicate the 
$x$-coordinate of the vortex pair barycenter at each loop transition.
In addition to the leapfrog dynamics, the data reveal a clear outward drift of the vortices. This drift arises from the gradual 
spreading of the background Gaussian beam, which is itself induced by the defocusing nonlinearity of the medium.
The vortex dynamics is theoretically reproduced in the bottom panel of Fig.~\ref{fig:setup}(c) which displays the results the point vortex model \cite{Helmholtz1867}. 
According to this model, the velocity of vortex $i$ of charge $\ell_i$ ($=\pm 1$) and position $\tilde{\boldsymbol{r}}_i(\tau)$ is
\begin{equation}
\boldsymbol{v}^{\rm{Pvm}}(\tilde{\boldsymbol{r}}_i) = 
\sum_{j\ne i}
\frac{\ell_j}{|\tilde{\boldsymbol{r}}_i - \tilde{\boldsymbol{r}}_j|^2}
\begin{pmatrix}
\tilde{y}_j - \tilde{y}_i \\
\tilde{x}_i - \tilde{x}_j
\end{pmatrix}
    + \boldsymbol{v}^{\rm{bgd}}(\tilde{\boldsymbol{r}}_i).
    \label{pvm}
\end{equation}
The point vortex simulation is performed by setting the same initial vortex configuration as in the experiment. The spreading of the beam is accounted for by including in Eq. \eqref{pvm} 
the background outward flow $\boldsymbol{v}^{\rm{bgd}}(\tilde{\boldsymbol{r}})$ measured in the absence of vortices (see Supplementary for details).
At later time, the vortex dynamics is dominated by the outward flow and the leapfrogging stops, as observed both in experiment and in the simulations  {(see Appendix D)}.\\

The data presented in Fig.~\ref{fig:setup} provide a clear observation of vortex leapfrogging in a quantum  fluid of light over more than one full cycle.
We now investigate the breakdown of leapfrogging and its robustness  with respect to initial conditions.
To do so, we vary the initial separation between the two vortex–antivortex pairs: the first pair is created at positions $(0,\pm 3\xi)$, a second one at positions $(5\xi,\pm 3\xi)$ leading to $\Delta R /\xi = 6$ and $\Delta r /\xi = 5$.
We then examine the resulting evolution over a single loop.
To characterize the topological structures of the flow, we compute the quantized circulation from the measured phase map, which allows us to identify the topological charge associated with each vortex.
In addition to vortex and anti-vortex, recent studies \citep{congy2024,Panico2025b} have highlighted the importance of other critical points, such as saddles and nodes characterized by a topological charge called the Poincaré index \citep{Strogatz2015} (see Appendix C for details).
Fig.~\ref{fig:vacuum}(a) shows the dynamics of the two vortex pairs and the evolution of critical points over half a leapfrogging cycle. 
The amplitude images (top panel) and the corresponding phase maps (bottom panel) 
% {\color{red}and the associated equiphase line (bottom panel)} 
are shown for $\tau_{l_0} = 89, 120, 123, 144$. 
The amplitude profile along the solid horizontal line is overlaid on each image.
The initial positions of the rightmost vortex at $\tau_{l_0} = 89$ is indicated with a dashed circle, indicating a first leapfrog at early time.
The phase maps are shown with the corresponding streamlines and critical points: vortices and antivortices are marked by blue and red dots, respectively, while saddle and node points are given as orange and green diamonds.

Between $89 < \tau_{l_0} < 120$, vortices with the same vorticity are separated by a saddle point. 
A low density area emerges between the $x>0$ vortex pair, inside which a new vortex-antivortex pair appears at $\tau_{l_0} = 120$ due to a pitching of the equiphase lines, accompanied by two saddles points.
The corresponding scenario of vortex formation has long been associated with phase slippage \cite{Langer1967}; it conserves both the vorticity and the Poincaré index as described in detail in Ref.~\cite{poincare1988}.
Due to a small asymmetry in the experiment, a new vortex-antivortex pair is created in the $y<0$ region. 
After the pair formation, the two vortices move apart (as described in \cite{Langer1967,poincare1988}) and the upper one (red dot) approaches $y=0$.
This is accompanied by an increase of the velocity (blue peak at $\tau=120$ in Fig.~\ref{fig:vacuum}(c)), immediately followed by a $2\pi$ phase jump (blue and teal curves in Fig.~\ref{fig:vacuum}(b)) and an inversion of the direction of the velocity when the axis has been crossed (teal dip at $\tau=123$ in Fig.~\ref{fig:vacuum}(c)). 
The instant when the vortex crosses the axis corresponds to a cancellation of the density, as seen in Fig.~\ref{fig:vacuum}(a) amplitude profiles.
Finally, at $\tau_{l_0} = 144$, the dynamics conclude with three vortex–antivortex pairs. The different steps of the whole process are detailed in the Supplemental Material.
It thus appears that the region of low intensity has acted as a phase slip center since this region shares many features with a (finite extent) dark soliton which, in our two dimensional configuration experiences snake-instability \cite{Zakharov1974,Kuznetsov1988} and decays by emitting vortex dipoles \cite{Law1993,Tikhonenko1996,Dutton2001}.\\

The complexity of the dynamics at this stage, and in particular the existence of an important compressible component of the velocity field, prohibits a direct test of the above interpretation using the feedback loop setup of Fig.~\ref{fig:setup}.
However, we performed another complementary experiment in which two counter-rotating multicharged vortices (with $\ell=\pm10$ quanta of circulation) are injected at positions $(0,\pm \tfrac12 \Delta\tilde{R})$. 
%%% FIG 3 %%%%
\begin{figure}[!t]
    \centering
    \includegraphics[width=0.99\linewidth]{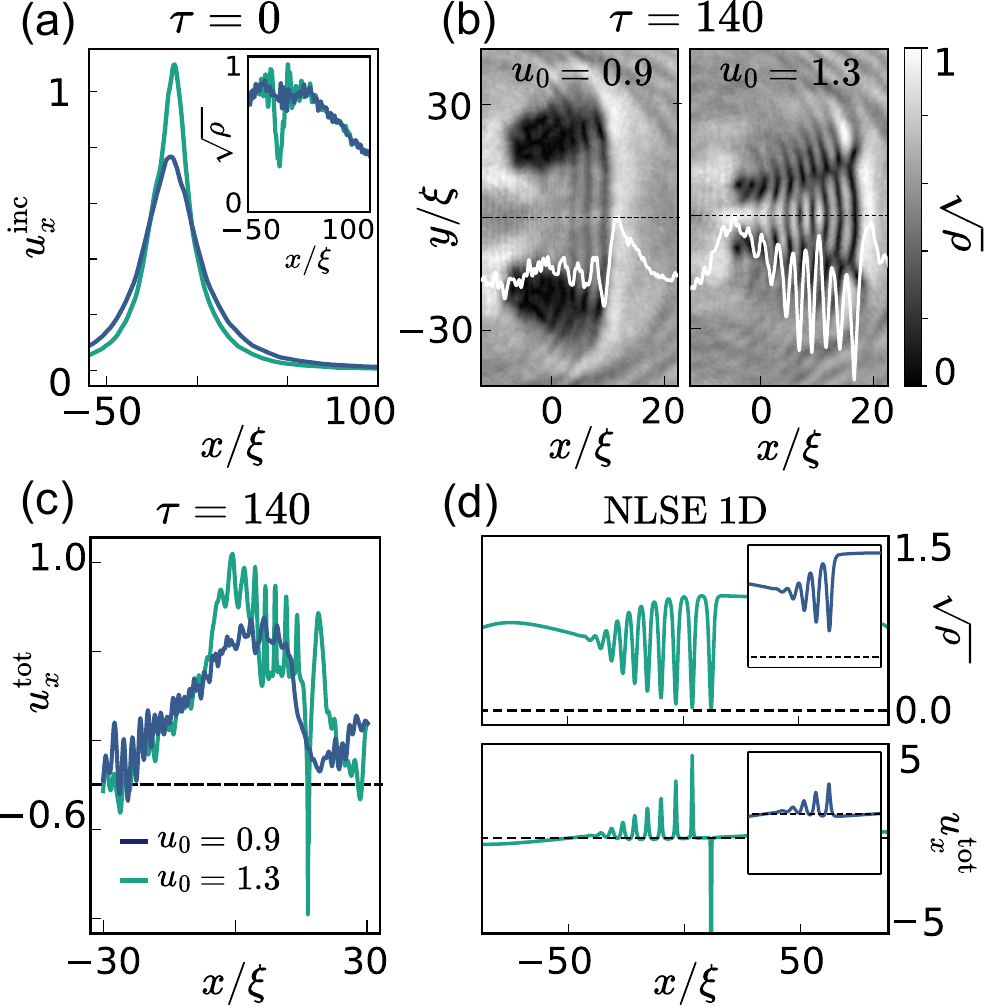}
    \caption{\textbf{Dispersive shock waves and wavetrain.} Two counter-rotating vortex of charge $\ell=\pm10$ are imprinted at the input of the medium with two separation distance of $\Delta\tilde{R}=43,30$.
    \textbf{(a)} - Input velocity profile along the x-axis between the vortices.
    The inset gives the associated amplitude profile.
    The blue and light blue curves shows the results for an injected velocity peak $u_0$ of respectively $0.9$ and $1.3$ in speed of sound unit.
    \textbf{(b)} - Output amplitude images at $\tau_{l_0}=140$ for the two configuration with their associated amplitude along the dark line.
    \textbf{(c)} - Total velocity along $x$. 
    \textbf{(d)} - Density (top), velocity (bottom) profile obtained in a one-dimensional NLSE with the two initial states shown in (a).}
    \label{fig:dsw}
\end{figure}
%%% FIG 3 %%%%
Fig.~\ref{fig:dsw} presents the results obtained for two different inter-vortex distances: $\Delta\tilde{R}=43$ and 30, with dark and light blue curves, respectively.
% The initial amplitude and velocity profiles along the horizontal axis between the vortices are shown in Fig.~\ref{fig:dsw}(a). 
Fig.~\ref{fig:dsw}(a) shows the initial (i.e., $\tau=0$)
longitudinal component $u_x^{\mathrm{inc}}$ of the incompressible part of the
velocity (see Appendix A) 
recorded
along the $x$-axis. The associated amplitude profile is presented in the inset.
In the present configuration a large horizontal velocity is initially imparted over an extended region. 
We measured the peak velocities in Fig.~\ref{fig:dsw}(a), $u_0\sim0.8$ and $u_0\sim1.1$ for $\Delta\tilde{R}=43$ and 30, respectively.
The expected maximum velocities from the point vortex model, are $u_0=4\ell/\Delta\tilde{R}=0.9$ and 1.3, respectively, in reasonable agreement with the experiment considering the strong inhomogeneity of the experimental density profile not present in the model.
Over a single time loop, the two multi-charged counter-rotating vortices propagate toward positive $x$ and split into $\ell$ singly charged vortices, as expected from decay of multi-charged vortices in superfluids \cite{Pethick2011}.
%\cite{pitaevskij_bose-einstein_2016}.
The singly charged vortices move approximately along straight trajectories, one for vortices of positive vorticity and another for those of negative vorticity, mainly guided by the self-induced velocity field \footnote{The point vortex model is not expected to be accurate in this configuration where the vortices are separated by distances of order $\xi$. However its use can help understanding the gross features of vortex motion: it would predict rapid oscillations of the vortices with small departure from a mean trajectory which indeed corresponds to the formation of a vortex street.}.
Due to the background outward flow, these trajectories are neither horizontal nor parallel but exhibit a drift away from the beam center, as seen in Fig.~\ref{fig:dsw}(b).

The vortex street of counter-rotating vortices channels a strong current localized in the region near the horizontal axis. To capture the flow behavior around 
$y\approx 0$, we employ an effective one-dimensional description and solve Eq.~\eqref{eq:NLSE} for $\psi(x,0)$ neglecting the $y$ dependence.
The initial density and velocity profiles used
in the simulation are matched to those extracted from the
experimental input data given in Fig.~\ref{fig:dsw}(a). 
This simplified description reveals that the stripes observed on the amplitude profile along the horizontal line in the region of the vortex street [see Fig.~\ref{fig:dsw}(b)] correspond to a dispersive shock wave (DSW). 
Indeed, the theoretical amplitude and total velocity profiles obtained in Fig.~\ref{fig:dsw}(d) display patterns similar to the ones measured along $y=0$, presented respectively in  Fig.~\ref{fig:dsw}(b) and (c): an upstream and a downstream smooth flows are separated by a region of nonlinear oscillations which constitutes the DSW.

A second argument supporting this interpretation is the difference in the behaviors between the two cases presented in Fig.~\ref{fig:dsw}. 
In the strong flow case ($\Delta\tilde{R}= 30$), at $\tau=140$ the envelope of the amplitude oscillations within the DSW cancels at $\tilde{x}_{\rm vp}\simeq18$, with an abrupt change of sign of the associated velocity profile at the same position. 
This is also observed in the numerical simulations of Fig.~\ref{fig:dsw}(d). 
Such a behavior is characteristic of the emergence of a "vacuum point" within the DSW \cite{El1995}. In this regime, the leading edge of the DSW propagates faster than the fluid ahead of it. Consequently, the segment of unperturbed fluid in front of the DSW shrinks when a vacuum point is present, whereas it expands when no vacuum point forms. 
The experimental data show indications of this trend and the effect is clearly and consistently observed in the numerical simulations.

This phenomenology confirms our initial hypothesis about the underlying one-dimensional dispersive hydrodynamic behavior of the fluid in the vortex-street region.
Moreover, the above discussion shows that, with these initial conditions, the dark soliton previously observed in the data corresponding to Fig.~\ref{fig:vacuum} is not only amplified as expected, but that it constitutes the leading edge of an entire DSW.
It is shown in the supplemental material \cite{supplemental} that the phase 
difference between the two ends of a such DSW is a continuously decreasing function of time. The discrete phase slip events previously observed in Fig. \ref{fig:vacuum} are now replaced by  the formation of a dispersive shock which constitutes a continuous and efficient source of dissipation.

In this work, we have experimentally investigated and characterized the temporal evolution of vortex leapfrogging in a quantum fluid of light.
Our platform offers control over the injected wavefunction and enables high-resolution measurements of the full field (amplitude {\it and} phase), allowing to track vortices evolving in an inhomogeneous background with an outward flow.
We compared the observed vortex dynamics with predictions from the point vortex model, and found good agreement in the leapfrogging regime.
However, by preparing initial configurations with strong inter-vortex currents, we observed the collapse of the leapfrogging mechanism associated with the occurrence of phase slip events.
Furthermore, we performed a complementary experiment in which two counter-rotating multicharged vortices were injected, reaching
a highly nonlinear configuration.
In this setting we observe the formation of a quasi one-dimensional dispersive shock wave. 
This DSW plays the role of a center of quantum dissipation constituting a continuous alternative to phase slip events.
Our work opens new perspectives for studying macroscopic vortex dynamics  present in vortex turbulence, through  microscopically controlled mechanisms, and more broadly contributes to the exploration of out-of-equilibrium quantum fluid physics and dissipation in superfluids.

\begin{acknowledgements}
The authors acknowledge insightful discussions with T. Aladjidi, T. Congy, P.-\'E. Larr\'e and Q. Schibler.
\end{acknowledgements}

\bibliography{main}

\clearpage
\onecolumngrid
\section*{End Matter}

\twocolumngrid
\textit{Appendix A: Velocity decomposition.} 

Direct access to the fluid phase, shown in Fig.~\ref{fig:setup}(b), allows a measurement of the velocity field, given by $\boldsymbol{v^{tot}}(\boldsymbol{r}) \propto \boldsymbol{\nabla}_{\perp}\phi(\boldsymbol{r})$ where $\phi$ is the phase of the order parameter (see Supplementary for details).
We introduce the density-weighted velocity, given by $\boldsymbol{u^{tot}(r)}=\sqrt{\rho(\boldsymbol{r})}\boldsymbol{v}^{tot}(\boldsymbol{r})$, where $\rho(\boldsymbol{r})$ is the light intensity.
We then identify the compressible and incompressible parts of $\boldsymbol{u^{tot}(r)}$ using the Helmholtz decomposition to separate the divergent (compressible) and rotational (incompressible) components:
\begin{equation}
    \boldsymbol{u^{tot}}(\boldsymbol{r})=
    \underbrace{\boldsymbol{\nabla}\phi(\boldsymbol{r})}_{\textrm{compressible}} + \underbrace{\boldsymbol{\nabla}\times\boldsymbol{A}(\boldsymbol{r})}_{\textrm{incompressible}}
\end{equation}
where $\phi$ is scalar and $\boldsymbol{A}$ a vector field. 
The same decomposition can be written in the Fourier space:
\begin{equation}
    \boldsymbol{U^{tot}}(\boldsymbol{k}) = i\boldsymbol{k}U_\phi(\boldsymbol{k})+i\boldsymbol{k}\times\boldsymbol{U_{A}}(\boldsymbol{k}),
\end{equation}
where 
\begin{equation}
U_\phi(\boldsymbol{k})=-i\frac{\boldsymbol{k}\cdot \boldsymbol{U^{tot}}(\boldsymbol{k})}{{k}^2}
, \quad
\boldsymbol{U_{A}}(\boldsymbol{k})=i\frac{\boldsymbol{k}\times \boldsymbol{U^{tot}}(\boldsymbol{k})}{{k}^2}.
\end{equation}
Thus, we write the definition of the compressible and incompressible part in the real space using the inverse Fourier Transform (TF):
\begin{equation}
    \begin{split}
        \boldsymbol{\nabla}\phi(\boldsymbol{r})&=\textrm{TF}^{-1}[i\boldsymbol{k}\cdot U_\phi(\boldsymbol{k})] \\
        \boldsymbol{\nabla}\times\boldsymbol{A}(\boldsymbol{r})&=\textrm{TF}^{-1}[i\boldsymbol{k}\times \boldsymbol{U_A}(\boldsymbol{k})].
    \end{split}
\end{equation}
We obtain the incompressible weighted velocity field by directly subtracting the compressible part from the total weighted velocity $\boldsymbol{u^{inc}}=\boldsymbol{u^{tot}}- \boldsymbol{\nabla}\phi(\boldsymbol{r})$.

\

\textit{Appendix B: Vortex wave function}.
If our experiment we initiate the dynamics by imprinting vortices in the laser beam through a SLM. The exact density profile of a vortex is not an analytic function, and it is more appropriate to use a realistic approximate form.
The order parameter describing 
a single charge vortex $\ell=\pm 1$ positioned at the origin is well described by the ansatz:
\begin{equation}
    \psi_v(\boldsymbol{r}) = \sqrt{\rho_0}\,
    \frac{\boldsymbol{r} \,\exp\{i\, \ell\, \theta\}}{\sqrt{{r}^2 + (\Lambda^{-1}\xi)^{2}}},
\label{vortex}
\end{equation}
where $\theta$ is the polar angle. The value $\Lambda=0.82$ was determined in Ref. \cite{bradley2012energy}
for enforcing matching the slope of the
ansatz density distribution to the exact value at the center
of the core. For implementing several vortices we use product of wave functions of type \eqref{vortex}, each centered at the desired 
position.

\

\textit{Appendix C: Topological charge}.
We use the quantization of the circulation to probe the topological charge in the system by numerically computing the circulation of the phase $\phi$ of the order parameter about round loops:
\begin{equation}
    I_{\rm\sss V} = \frac{1}{2\pi}\oint d\phi = 0,\pm 1, \cdots
\end{equation}
$I_{\rm\sss V}$ is the vorticity, a topological charge associated to each vortex.

\begin{figure}[h!]
    \centering
    \includegraphics[width=0.99\columnwidth]{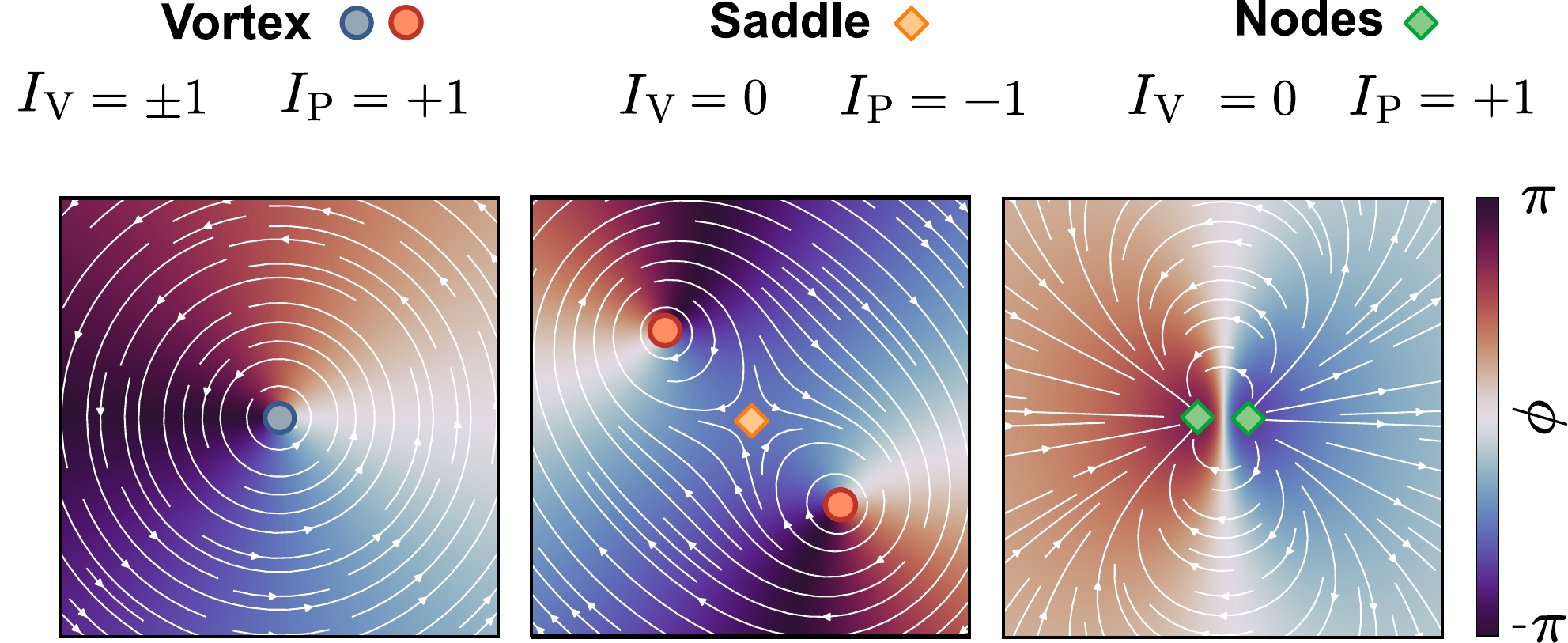}
    \caption{\textbf{Critical points} - From left to right, phase map of a positive vortex, an anti-vortex pair and two phase extrema, respectively.
    The associated streamlines are superimposed on the images.
    Vortex and anti-vortex are represented by blue and red dots, saddle and nodes by orange and green diamond.
    }
    \label{fig:topo}
\end{figure}

Recent studies \cite{congy2024, Panico2025b} have underlined the relevance of another topological index, called the Poincar\'e 
index, which is associated to the winding of the direction of the flow velocity around close loops. It is defined as
\begin{equation}
I_{\rm\sss P} = \frac{1}{2\pi}\oint d\theta_v= 0,\pm 1, \cdots
\end{equation}
where $\theta_v$ is the polar angle of $\boldsymbol{v^{tot}}$, given by $\theta_v = \atan (v_y/v_x)$.
Vortices carry a Poincar\'e index +1 (irrespective of their vorticity). Besides vortices, other critical points carry a finite Poincar\'e index: these are zeros of the velocity field, 
namely saddles (saddle points of the phase $\phi$) and 
nodes (extrema of $\phi$).
All these critical points are schematically represented 
in Fig. \ref{fig:topo}. \\

\

\begin{figure}[h]
    \centering
    \includegraphics[width=0.99\columnwidth]{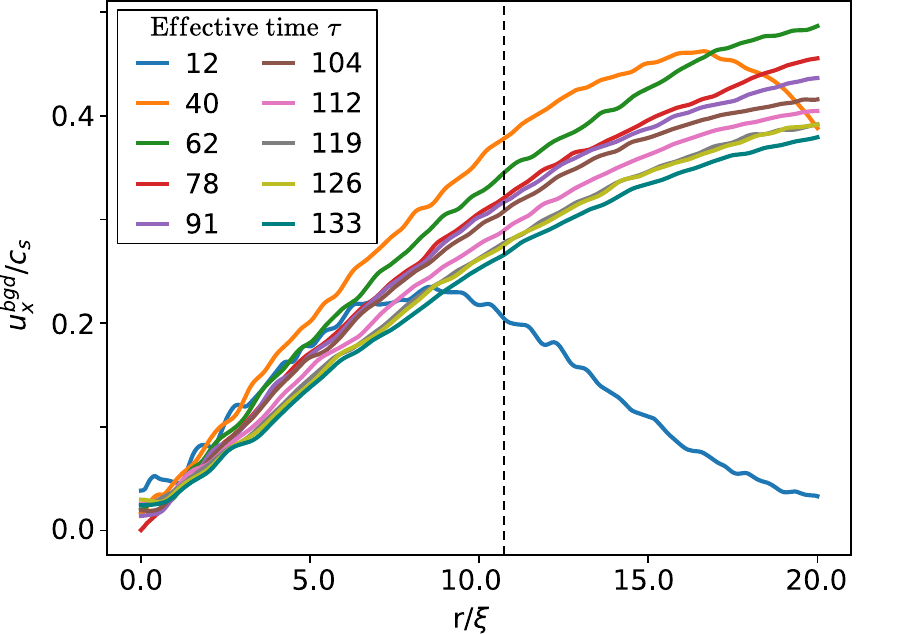}
    \caption{\textbf{Background velocity} - Measured outward flow versus $r/\xi$ from the beam center.
    Each color plot gives the background velocity at different value of $\tau$ in a single loop.
    The dashed line at $r/\xi\sim11$ indicate the vortex pair barycenter at the last loop of Fig.~\ref{fig:setup}(c).}
    \label{fig:v_bgd}
\end{figure}

\textit{Appendix D: Fluid outward flow}.
We assume that the vortices remain near the center of the beam’s Gaussian intensity profile during the loop transitions indicated by the dashed lines in Fig.~\ref{fig:setup}(c).
As shown in Fig.~\ref{fig:v_bgd}, for radial distances $r/\xi < 11$, the outward flow remains nearly identical across all values of $\tau$.
This justifies our assumption that the vortex dynamics occur in a quasi-stationary background during the early stages of evolution.
However, due to the progressive displacement of the vortex pairs’ barycenter, the vortices eventually move beyond $r/\xi\sim11$, where the outward flow profile changes significantly.
As a result, no more than four complete leapfrogging loops can be realized experimentally, since beyond this point, the background conditions are no longer consistent across successive cycles.

\end{document}

\author{Myrann Baker-Rasooli~\orcidlink{0000-0003-0969-6705}}
\affiliation{Laboratoire Kastler Brossel, Sorbonne Universit\'e, CNRS,
ENS-PSL Research University, Coll\`ege de France, 4 Place Jussieu, 75005 Paris, France}

\author{Nathan du Toit~\orcidlink{0009-0000-4988-4722}}
\affiliation{Université Paris-Saclay, CNRS, LPTMS, 91405 Orsay, France}

\author{Nicolas Pavloff~\orcidlink{0000-0001-8814-7848}}
\affiliation{Université Paris-Saclay, CNRS, LPTMS, 91405 Orsay, France}
\affiliation{Institut Universitaire de France (IUF)}

\author{Quentin Glorieux~\orcidlink{0000-0003-0903-0233}}
\affiliation{Laboratoire Kastler Brossel, Sorbonne Universit\'e, CNRS,
ENS-PSL Research University, Coll\`ege de France, 4 Place Jussieu, 75005 Paris, France}
\affiliation{Institut Universitaire de France (IUF)}

\title{Supplementary Material.\\
Vortex leapfrogging and superfluid dissipation mechanisms in a  fluid of light}
\maketitle

\section*{Experiment details}

\begin{figure}[hb!]
    \centering
    \includegraphics[width=0.95\columnwidth]{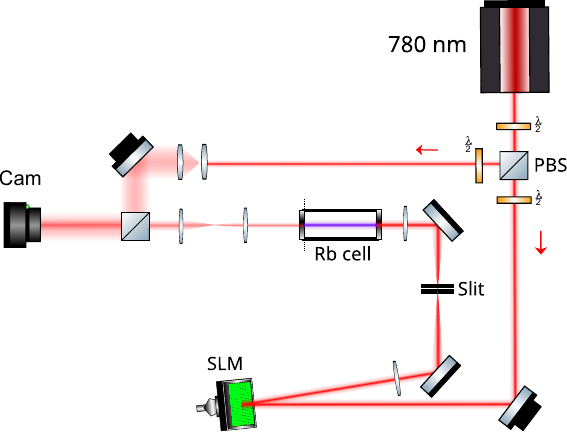}
    \caption{\textbf{Experimental Setup Detail} - A 780 nm laser beam is directed onto a Spatial Light Modulator (SLM), which is imaged at the input of a 20 cm-long rubidium vapor cell. The phase modulation produces four rotating vortices in the transverse plane of the beam. The output plane of the nonlinear medium is then imaged on a camera. A reference beam, separated from the initial laser beam, is recombined with the main beam before the camera to enable phase measurement.}
    \label{fig:manip_detail}
\end{figure}

To create the fluid of light, we use a continuous-wave diode laser at 780 nm, with tunable detuning relative to the $^{87}$Rb D2 resonance line, as illustrated in Fig. \ref{fig:manip_detail}. 
A $\pm 2\pi$ phase circulation pattern is applied to the Spatial Light Modulator (SLM) to generate the dipole vortex.
To eliminate unmodulated reflections from the SLM, a vertical grating is superimposed on the horizontal one, and only the first vertical order is selected in the Fourier plane using a slit.
The image from the SLM is then relayed to the entrance of the nonlinear medium via a telescope, with the beam waist at the entrance measuring $\omega = 1.7$ mm.
The nonlinear medium consists of a 20 cm rubidium cell containing a pure state of $^{87}$Rb.
The output plane of the cell is imaged onto a camera without protective glass. 
Using a reference beam separated from the main beam upstream, we access the phase of the two counter-propagating beams by analyzing the interference fringes formed after recombination before the camera.
To adjust the effective interaction time $\tau$, the laser power is controlled using a rotating $\lambda/2$ wave plate, with a maximum power of 5W. 
During this process, the cell temperature is maintained at 150°C, and the laser detuning is fixed at $\Delta = -6$~GHz relative to the $^{87}$Rb $F = 2 \to F^{'}=2$ transition.

\section*{Vortices time evolution}

\begin{figure}[!b]
    \centering
    \includegraphics[width=0.95\columnwidth]{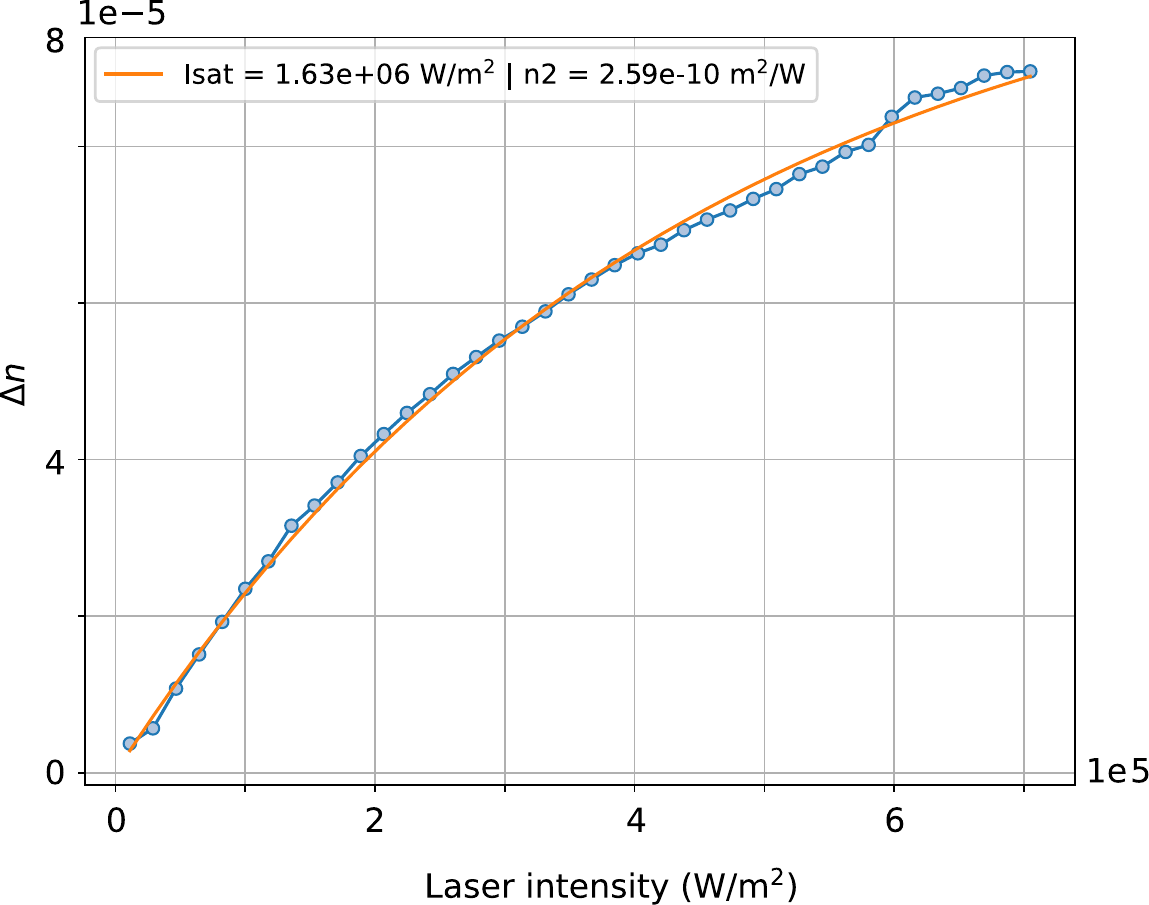}
    \caption{\textbf{Non-linear refractive index accumulated along the medium} - Each point is measured for a given beam intensity and the data are fitted from $n_2 \frac{I}{1+I/Is}$ (with $I_s$ the saturation intensity) given by the orange curve.}
    \label{fig:non-linear phase}
\end{figure}

To observe the time evolution of the dipole vortex, we vary the laser power and measure the fluid without the vortex (Gaussian profile) to determine the nonlinear phase shift $\Delta \phi$ accumulated within the cell. 
This phase shift is linked to the nonlinear refractive index through the formula $\Delta n = \Delta\phi / (k_0 L)$, as shown in Fig. \ref{fig:non-linear phase}.
This measurement allows us to obtain the value of $\tau = \Delta\phi$, and consequently, the parameter $\Bar{\xi} = \sqrt{L/(k_0 \tau)}$.
Next, we imprint the two phase circulations separated by a distance $\Delta \boldsymbol{r}$ while maintaining the ratio $\Delta \boldsymbol{r}/\Bar{\xi}$ constant.
The barycenter position of the two-phase circulation from the center of the beam $\boldsymbol{r}_0$ is also adjusted in order to keep the quantity 
$\boldsymbol{r}_0/\Bar{\xi}$ constant.

\begin{figure}[!t]
    \centering
    \includegraphics[width=0.99\columnwidth]{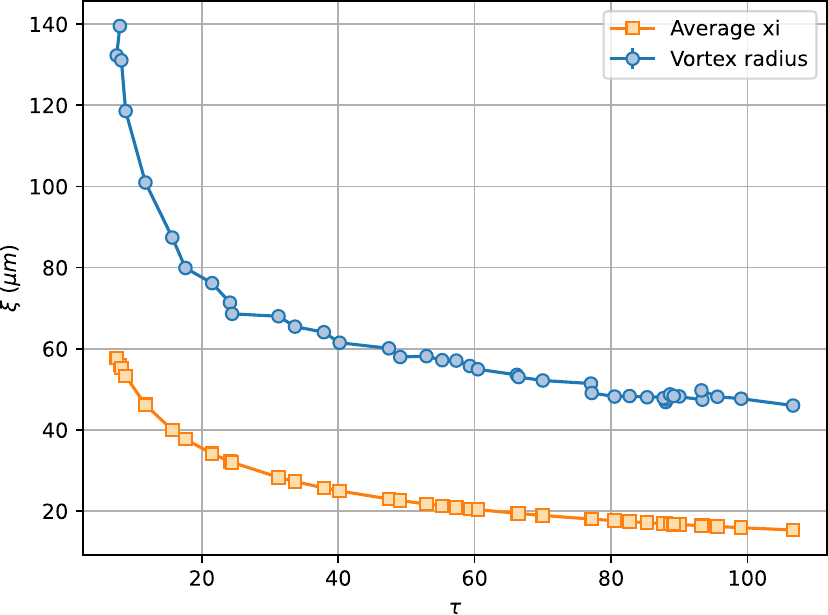}
    \caption{\textbf{Measurement of the final healing length} - Each point is measured for a given beam intensity and the data are fitted from the vortex core amplitude.}
    \label{fig:xi_vortex}
\end{figure}

In addition to measuring the fluid with and without the vortices, we also measure the final value of $\xi$ for each $\tau$ shown in Fig.\ref{fig:xi_vortex} by injecting a single vortex and determining its radius through its radial amplitude profile.

\section*{Velocity map reconstruction}

\begin{figure}[h]
    \centering
    \includegraphics[width=0.99\columnwidth]{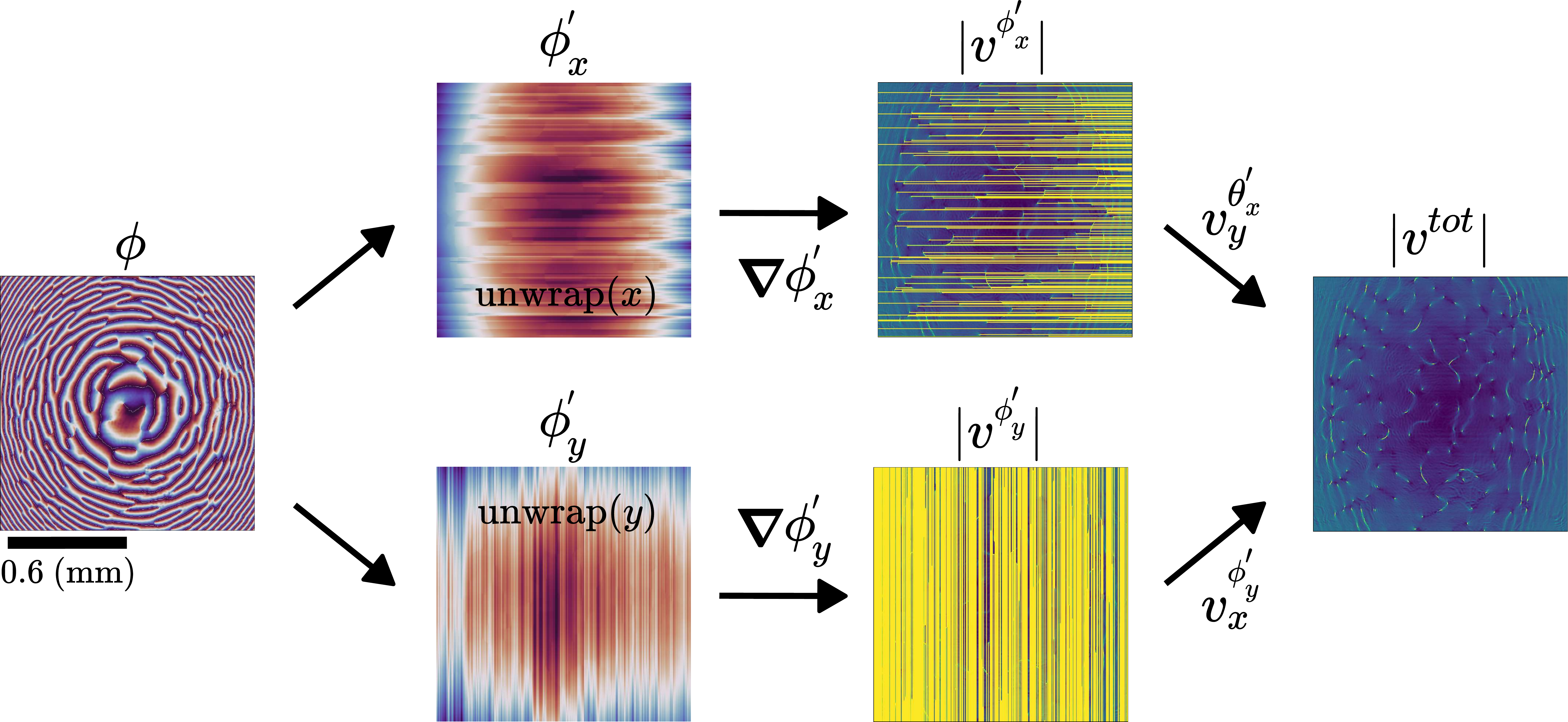}
    \caption{\textbf{Total velocity reconstruction} - The total velocity is the combination of each gradient component calculated from the 1D unwrapped phase.}
    \label{fig:velo_tot}
\end{figure}

From the superfluid phase map $\phi=\phi(\boldsymbol{r}_{\perp})$, we reconstruct the total velocity field of the flow, ${v}(\boldsymbol{r}_{\perp})=\nabla_{\perp}\phi(\boldsymbol{r}_{\perp})/k_{0}\equiv\boldsymbol{k}_{\perp}$.
To compute the velocity map, we must first perform a 2D unwrapping of the phase.
However, because we work with discretized data arrays, phase singularities (i.e., $2\pi$ windings) create numerical discontinuities that prevent an accurate reconstruction.
To circumvent this issue, the phase is unwrapped separately along each axis, yielding $\phi_{x}'$ and $\phi_{y}'$.
The total velocity field is then obtained by combining the gradient components computed from these two independently unwrapped phase profiles, as illustrated in Fig.~\ref{fig:velo_tot}.

\section*{Bifurcations of the flow pattern}

\begin{figure*}
\includegraphics[width=0.96\linewidth]{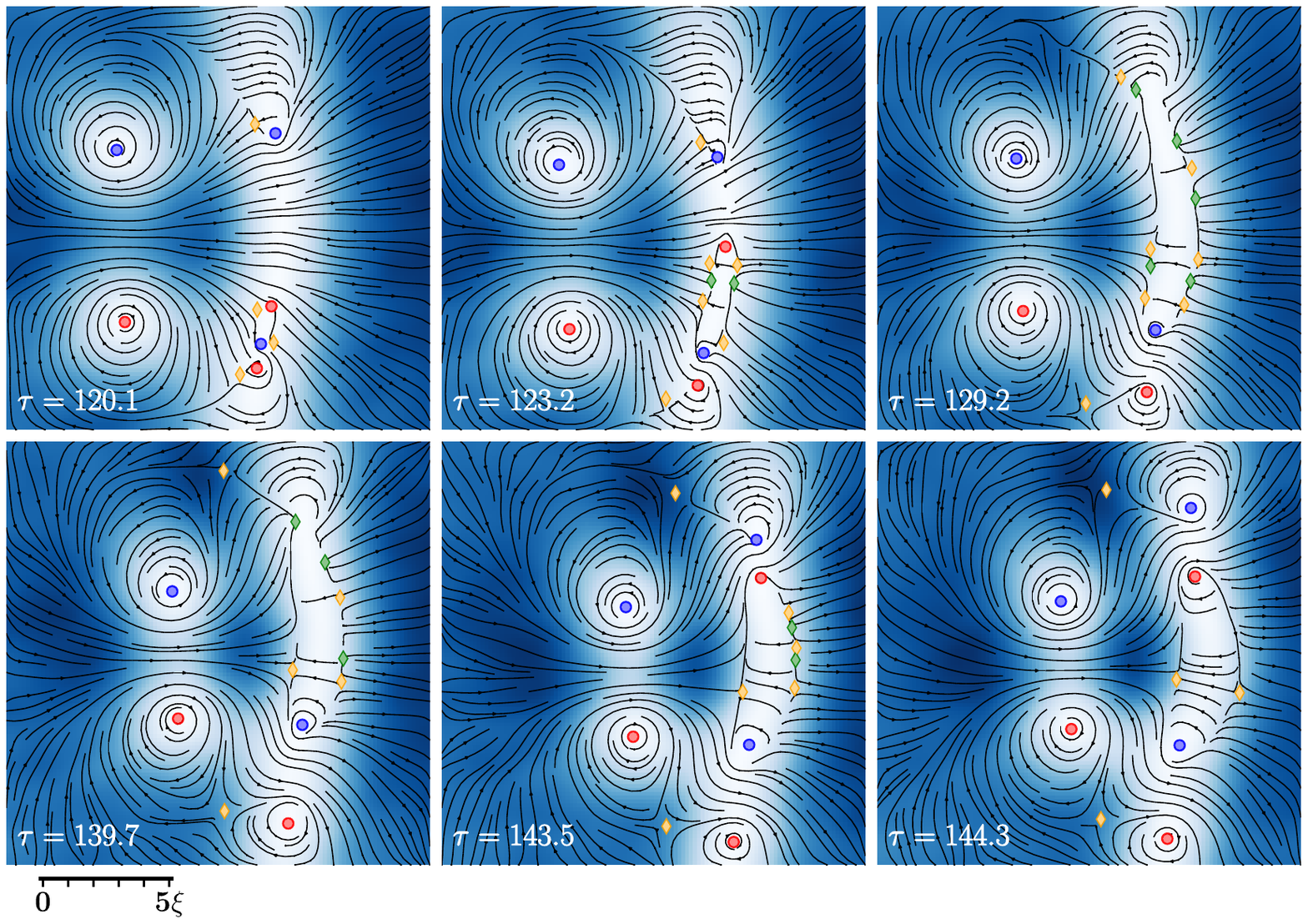}
\caption{Evolution of the flow pattern displayed in Fig. 2 of the main text. The different panels are labeled with their respective times $\tau$. Dark blue regions correspond to larger intensity of light. Positive (negative) vortices appear as blue (red) dots. Nodes and saddles are represented as green dots and yellow diamonds, respectively. The oriented black solid curves are streamlines. The experimental data have been smoothed over a 3 pixels mesh size, small compared to the typical healing length (8 pixels).}\label{fig.bifur}
\end{figure*}

In this section we illustrate how critical points 
may appear and disappear during the dynamical evolution of the flow. 
We consider the flow depicted in Fig. 2 of the main text during the time laps
$100 < \tau < 145$. In the initial stage, the system displays two pairs 
of vortices and two saddles, both located between vortices of the same vorticity. At time $\tau=120.1$ a new pair of vortices appears (close to the lower right negative vortex in Fig. \ref{fig.bifur}), along
with two new saddles, as described by the so-called NHH scenario 
\cite{poincare1988} which corresponds to a fold-Hopf bifurcation \cite{congy2024}. 
Then two supercritical pitchford bifurcations occur, during which the recently created 
saddles both give birth to a new saddle and a new node ($\tau=123.2$ in Fig. \ref{fig.bifur}). This is connected to a drifting apart of the two new vortices, one of which (the upper one) collides with the upper right positive vortex. The two vortices involved in this collision annihilate and give birth to two nodes (another type of fold-Hopf bifurcation, also characterized 
in Ref. \cite{congy2024}). The subsequent occurrence of a saddle-node bifurcation (in the right part of the upper half plane) yields the flow pattern 
observed at $\tau=129.2$ in Fig. \ref{fig.bifur}. From this moment on, the flow pattern simplifies in the lower half plane: two 
supercritical pitchforks each eliminate a saddle and a node close to the lowest positive vortex, resulting in the pattern observed at $\tau=139.7$. Then, the collision of two nodes in the upper half plane gives rise to two vortices of opposite sign (this is the mechanism opposite to that observed between times $\tau=123.2$
and 129.2). This process is followed by the appearance of a saddle-node, as can be observed at $\tau=143.5$ in 
Fig. \ref{fig.bifur}. Finally, two saddles and two nodes disappear, 
resulting in the flow observed at 
$\tau=144.3$. 

The above detailed description makes it apparent that, despite the fact that  the flow pattern may display complicated and rapidly evolving structures, at each step the topological constraints are duly respected. For example, the vorticity and Poincar\'e index in all the panels of Fig. \ref{fig.bifur} keep constant values $I_{\rm\sss V}=0$ and $I_{\rm\sss P}=2$. Note also that the elementary bifurcations involved during the evolution of the flow pattern have been previously observed in different settings \cite{Karman1998,Soskin1999,congy2024,Panico2025b}, and it has been shown in Ref. \cite{congy2024} that they are valid in any generic quantum fluid.

\section*{Phase shift across a dispersive shock wave}

\begin{figure}
\includegraphics[width=0.99\linewidth]{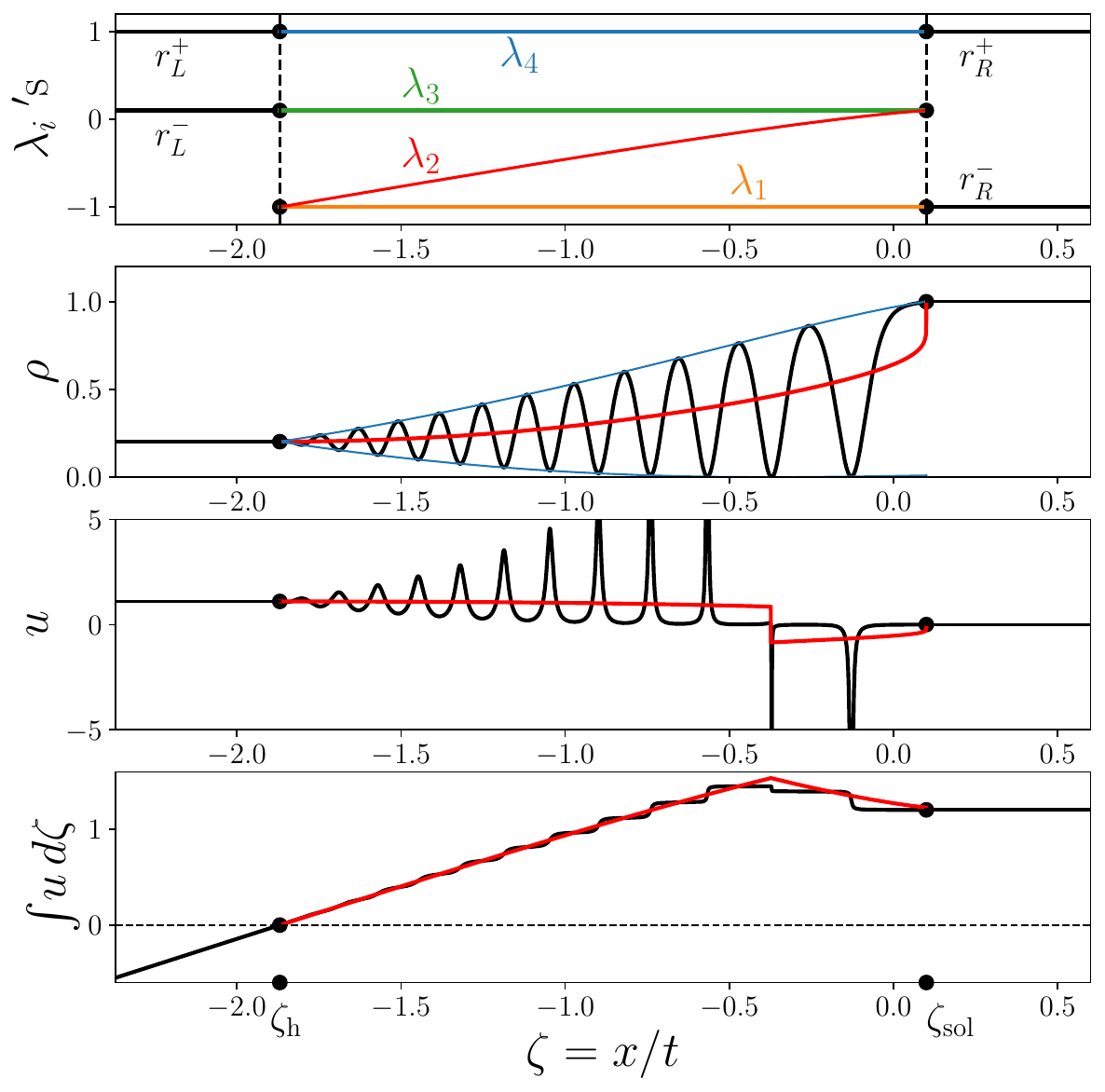}
\caption{Plot of characteristic quantities for a DSW whith
$\rho_R=1$, $u_R=0$, $u_L=1.1$ and $\rho_L=0.2$. In each 
panel the black circles mark the harmonic and solitonic boundary of the 
DSW, at $\zeta_{\rm h}=-1.9$ and $\zeta_{\rm sol}=0.1$, respectively. 
The upper plot displays the arrangement of the Riemann invariants 
specific to the DSW we consider.
In the three lower panels the 
red solid lines display the average quantities, $\bar{\rho}(\zeta)$, 
$\bar{u}(\zeta)$  
and $\int\!\bar{u}\,{\rm d}\zeta$, while the black solid lines are
$\rho(x,t)$, $u(x,t)$ and $\int\! u \,{\rm d}\zeta$ as 
computed from \eqref{dsw1}. The discontinuity of $\bar{u}$ at the vacuum point is $\bar{u}|_{\rm v.p.}^- - 
\bar{u}|_{\rm v.p.}^+
= -\pi (\lambda_3-\lambda_1)/K(m)$ where all the quantities are to be evaluated at the position $\zeta_{\rm v.p.}$ of the vacuum point, i.e., when $\lambda_2=\lambda_2|_{\rm v.p.}=\lambda_4-\lambda_3+\lambda_1$.
The two blue curves display the upper and lower envelopes of the density oscillations, $\rho_0(\zeta)$ and $\rho_0(\zeta)+A(\zeta)$, respectively.
}\label{fig.DSW1}
\end{figure}

In this section we determine the time dependence of the difference in phase of the order parameter between the two asymptotic ends of a dispersive shock wave (DSW). The computation can be conducted analytically in the case of a perfect Gurevitch-Pitaevskii configuration corresponding to piece-wise Riemann initial conditions \cite{El1995}. In such a case, the large scale features of the shock wave depend only on the self-similar quantity $\zeta=x/t$ and
the shock is characterized by Riemann invariants which behave as sketched in Fig.  \ref{fig.DSW1}.
The shock propagates to the right, with a leading edge of solitonic type, and a harmonic trailing edge. We denote the asymptotic velocities and densities as $u_L$ and $\rho_L$ (in the plateau region at the left of the DSW) and $u_R$ and $\rho_R$ (in the plateau at the right) and the dispersionless 
Riemann invariants are the  quantities \footnote{We use here reduced units as done in Equation (2) of the main text.} 
$r_R^\pm=\tfrac12 u_R \pm \sqrt{\rho_R}$
and $r_L^\pm=\tfrac12 u_L \pm \sqrt{\rho_L}$. They verify $r_R^+=r_L^+\ge r_L^-\ge r_R^-$. The density and velocity within the DSW read (see e.g., \cite{Kamchatnov-book,El2016})
\begin{equation}\label{dsw1}
\rho(x,t)=\rho_0+A\sn^2(W,m),\quad u(x,t)=V + C/\rho(x,t),
\end{equation}
where $\sn$ is the Jacobi elliptic sine function and
\begin{equation}
\begin{split}
\rho_0=& \tfrac14 (\lambda_4-\lambda_3-\lambda_2+\lambda_1)^2, \quad
A=(\lambda_4-\lambda_3)(\lambda_2-\lambda_1),\\
V=& \tfrac12 (\lambda_1+\lambda_2+\lambda_3+\lambda_4),\quad
m=\frac{(\lambda_2-\lambda_1)(\lambda_4-\lambda_3)}{(\lambda_4-\lambda_2)(\lambda_3-\lambda_1)},\\
W=& \sqrt{(\lambda_4-\lambda_2)(\lambda_3-\lambda_1)}(x-V t),\\
C=& \tfrac18 
(\lambda_3+\lambda_4-\lambda_1-\lambda_2)
(\lambda_2+\lambda_4-\lambda_1-\lambda_3)\times\\
& (\lambda_1+\lambda_4-\lambda_2-\lambda_3).
\end{split}
\end{equation}
In the above expressions $\lambda_4\ge \lambda_3\ge \lambda_2\ge \lambda_1$ are the Riemann invariants of the modulation equations.
Three of them are constant, they
are given by $\lambda_4=r_L^+=r_R^+$, $\lambda_3=r_L^-$ and $\lambda_1=r_R^-$, see Fig. \ref{fig.DSW1}. $\lambda_2$ is a function of $\zeta$ defined by the implicit equation
\begin{equation}
    \zeta=V+\frac{(\lambda_3-\lambda_2)(\lambda_2-\lambda_1)K(m)}
    {(\lambda_3-\lambda_2)K(m)-(\lambda_3-\lambda_1)E(m)},
\end{equation}
where $K$ and $E$ are the complete elliptic integral of the first and second kind, respectively. 
The boundary 
of the leading edge of the DSW (solitonic edge) is the quantity
\begin{equation}
    \zeta_{\rm sol}=\tfrac12 (\lambda_1+2\lambda_3+\lambda_4),
\end{equation}
while for the harmonic trailing edge
\begin{equation}
    \zeta_{\rm h}=2\lambda_1+\frac{1}{2} 
    \frac{(\lambda_4-\lambda_3)^2}{\lambda_3+\lambda_4-2\lambda_1}.
\end{equation}
The behavior of $\rho$ and $u$ is displayed in Fig \ref{fig.DSW1}.
This figure corresponds to the case $u_R=0$, $\rho_R=1$ (a configuration that can always be obtained through a suitable change of reference frame and a rescaling of
the density)
and $\rho_L=0.2$. Since 
$r_L^+=r_R^+$ ($=1$ here) this fixes $u_L=2(1-\sqrt{\rho_L})=1.1$. We are here in a configuration in which $\rho_L<\tfrac14 \rho_R$ and a vacuum point is formed in the DSW \cite{El1995}, as can be seen in the two central panels. The vacuum point corresponds to a zero of the lower envelope of the density pattern, i.e., of the quantity $\rho_0$. This occurs when $\lambda_2=\lambda_4-\lambda_3+\lambda_1=2\sqrt{\rho_L}+\tfrac12 u_R -\sqrt{\rho_R}=-0.1$.

We want to compute the time-dependent phase shift across the DSW. In our quantum fluid the velocity $u(x,t)$ is the gradient of the phase $\phi(x,t)$ of the order parameter, and we thus need to evaluate the quantity
\begin{equation}
    \Delta \phi(t)=\phi(x_{\rm max},t)-\phi(x_{\rm min},t)
    =\int_{x_{\rm min}}^{x_{\rm max}}\!\!\! u(x,t)\, {\rm d}x,
\end{equation}
where
$x_{\rm min}$ and $x_{\rm max}$ are positioned far away in the asymptotic plateau to the left and to the right, respectively. This integral can be evaluated by separating the integration domain in 3 regions: a portion at the left of the DSW ($x_{\rm min} < x<t\,\zeta_{\rm h}$) where $u$ takes the constant value $u_L$, another one at its right
($t\,\zeta_{\rm sol} < x<x_{\rm max}$ with $u(x,t)=u_R$), and finally the region of the DSW, with $t\,\zeta_{\rm h} < x<t\,\zeta_{\rm sol}$ in which the expression of $u$ is given by \eqref{dsw1}. To perform the integration in the region of the DSW we approximate the velocity field $u(x,t)$ by its spatial average $\bar{u}(\zeta)$:
\begin{equation}
\begin{split}
\bar{u}(\zeta) & =\frac{1}{2 K(m)}
   \int_0^{2 K(m)}\!\!\!\! u(x,t) \, {\rm d}x\\ 
&   = V +\frac{C}{\rho_0\, K(m)}\, 
\Pi\left(-\frac{A}{\rho_0},m\right),
\end{split}
\end{equation}
where $\Pi$ is the complete elliptic function of the third kind. 
Although this approximation performs poorly when used to directly evaluate $u(x,t)$, 
it proves quite effective for computing its integral, as illustrated in
Fig. \ref{fig.DSW1}. The approximation further improves 
at large time, when the DSW contains many oscillations.
This leads to the large time expression of $\Delta\phi$:
\begin{equation}\label{dsw8}
    \Delta \phi(t) = \Delta\phi(0) + \Gamma \, t,
\end{equation}
where 
\begin{figure}
\includegraphics[width=0.99\linewidth]{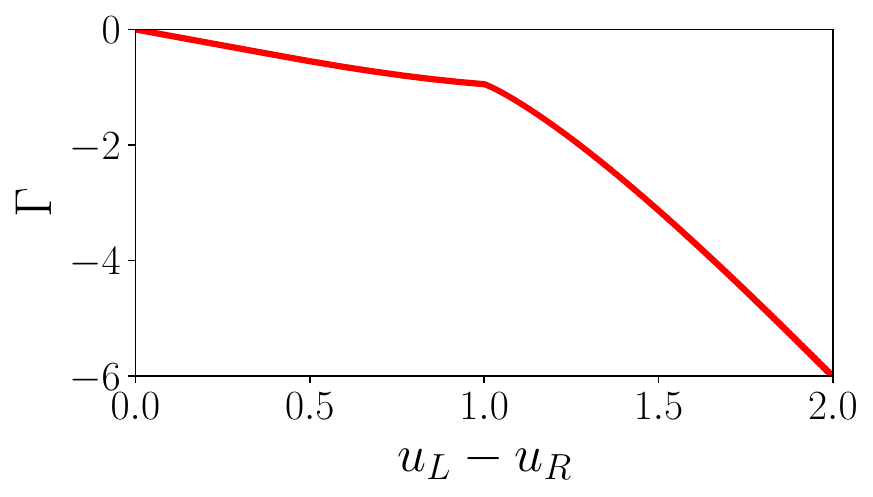}
\caption{Rate $\Gamma$ of change of the phase difference $\Delta\phi$ between the two ends of a DSW plotted as a function of the velocity difference $u_L-u_R$ for a configuration with $\rho_R=1$ and $u_R=0$.  
$\Gamma$ is computed by numerical integration of expression \eqref{dsw9}.
When $u_L=u_R$ there is no DSW and $\Gamma=0$. When $u_L-u_R=2$ the density $\rho_L$ cancels. 
The kink in the curve at $u_L-u_R=1$ corresponds to the formation of a vacuum point within the DSW (which exists for all DSWs with $u_L-u_R\ge 1$). The plots of Fig. \ref{fig.DSW1} correspond to the special case $u_L-u_R=1.1$.}\label{fig.DSW2}
\end{figure}

\begin{equation}\label{dsw9}
\begin{aligned}
\Delta\phi(0) &= u_R\, x_{\max} - u_L\, x_{\min},\\[6pt]
\text {and}\quad  \Gamma &= u_L\, \zeta_{\rm h}
        - u_R\, \zeta_{\rm sol}
+ \int_{\zeta_{\rm h}}^{\zeta_{\rm sol}} \!\!\!\bar{u}(\zeta) \, \mathrm{d}\zeta .
\end{aligned}
\end{equation}
The quantity $\Gamma$ is independent on $x_{\rm max}$ and 
$x_{\rm min}$. It
is the rate at which the phase of the order parameter changes between the two asymptotic ends of the DSW. It is represented as a function of $u_L-u_R$ in Fig. \ref{fig.DSW2}. We see that $\Gamma$ is always negative, indicating that the existence of the DSW is associated to a phase decreasing at constant rate
across the sample: the DSW acts as a continuous source of quantum dissipation in the superfluid.

\bibliography{main}